# *In vivo* recording of aerodynamic force with an aerodynamic force platform

David Lentink, Andreas F. Haselsteiner and Rivers Ingersoll
Department of Mechanical Engineering, Stanford University, Stanford, USA.

**Flapping wings enable flying animals and biomimetic robots to generate elevated aerodynamic forces. Measurements that demonstrate this capability are based on tethered experiments with robots and animals, and indirect force calculations based on measured kinematics or airflow during free flight. Remarkably, there exists no method to measure these forces directly during free flight. Such *in vivo* recordings in freely behaving animals are essential to better understand the precise aerodynamic function of their flapping wings, in particular during the downstroke versus upstroke. Here we demonstrate a new aerodynamic force platform (AFP) for nonintrusive aerodynamic force measurement in freely flying animals and robots. The platform encloses the animal or object that generates fluid force with a physical control surface, which mechanically integrates the net aerodynamic force that is transferred to the earth. Using a straightforward analytical solution of the Navier-Stokes equation, we verified that the method is accurate. We subsequently validated the method with a quadcopter that is suspended in the AFP and generates unsteady thrust profiles. These independent measurements confirm that the AFP is indeed accurate. We demonstrate the effectiveness of the AFP by studying aerodynamic weight support of a freely flying bird *in vivo*, which demonstrates that its upstroke is inactive.**

1. Introduction

The current method to measure the aerodynamic force of a flapping wing directly; is to tether an animal or robot and measure the forces transferred through the tether with a load cell (*1-3*). Tethered experiments with animals have many obvious concerns, but even tethered robot experiments are inaccurate when confounding inertia forces cannot be accounted for through dynamic modeling or measurement (*4*). Whereas tethered experiments have been the primary solution for evaluating aerodynamic force; measurements during free flight maneuvers are intrinsically more informative. During free animal movement, aerodynamic force can be calculated nonintrusively in two ways. First, the body mass and acceleration distribution can be measured and integrated using dynamics models to calculate force; this method requires sacrificing animals after their body kinematics has been measured (*5, 6*). Alternatively, the airflow can be measured around the animal and integrated using (simplified versions of) the Navier-Stokes equations to calculate the net aerodynamic force (*7-15*). Both calculations are based on indirect measurements of variables that need to be differentiated and integrated, which greatly amplify the effects of spatial and temporal measurement errors. In addition, these methods are based on high-speed video recordings that require massive data storage, and are computationally expensive to process, which can only be resolved by off-line computation of relatively small datasets. A nonintrusive, real-time, direct force measurement method (similar to the instrumented tether) does not exist for studying free locomotion in fluids. For studying terrestrial locomotion such a solution does exist: the force platform, e.g. (*16*). Here we present an aerodynamic force platform (AFP) that enables such measurements in fluids. We first establish the new method with an analysis using the Navier-Stokes equations, then validate it with a tethered quadcopter, and finally we demonstrate *in vivo* recordings for a freely flying bird.

## 2. Theoretical analysis

The aerodynamic force platform (AFP) is a box, instrumented with load cells, that encloses the object or animal that generates the net unsteady fluid force, figure 1*a*. It works based on Newton's third law applied to a fluid; the unsteady net fluid force needs to be supported by an equal and opposite net force that acts on the control volume boundary. The AFP is thus a mechanical representation of the control surface integral of the Navier-Stokes equation (*17*) that calculates the net unsteady force:

$$\bar{F}(t) = -\rho\frac{\partial}{\partial t}\underbrace{\iint_{CS}\bar{x}(\bar{u}\cdot\bar{x})dS}_{unsteady} - \rho\underbrace{\iint_{CS}\bar{u}(\bar{u}\cdot\bar{n})dS}_{convective} - \underbrace{\iint_{CS}p\bar{n}dS}_{pressure} + \underbrace{\iint_{CS}(\bar{\bar{\tau}}\cdot\bar{n})dS}_{shear} \quad (1.1)$$

In which $\rho$ is density, $\bar{x}$ is position, $\bar{u}$ is velocity, $dS$ is the integration surface, $\bar{n}$ is the surface normal vector, $p$ is pressure, and $\bar{\bar{\tau}}$ is the shear stress tensor. This net fluid force can be integrated exactly, provided the three-dimensional velocity, the pressure, and shear stress are known over the complete control surface as a function of time. In addition, small phase differences due to the finite propagation speed of pressure waves (sound) must be small (*18*), which is the case when the "AFP number" of the control volume is much smaller than one:

$$AFP_n = \frac{Lf}{a} \ll 1 \quad (1.2)$$

This condition is met when the control volume has an order of magnitude smaller length scale, $L$, than the distance sound travels, at speed $a$, within the shortest period of interest ($1/f$; in which $f$ is the frequency) that needs to be resolved in the fluid force $\bar{F}$. Under this condition the control surface integral (1.1) can be accurately evaluated. The contour integral is simplified by substituting the no-slip and no-flow condition on the surface of the physical control surface:

$$\bar{u}(\bar{x},t) = \bar{0} \quad (1.3)$$

This gives the following surface integral for the net fluid-dynamic force, which depends on the pressure and shear stress distribution that acts on the surface of the aerodynamic force platform:

$$\bar{F}(t) = -\underbrace{\iint_{CS}p\bar{n}dS}_{pressure} + \underbrace{\iint_{CS}(\bar{\bar{\tau}}\cdot\bar{n})dS}_{shear} \quad (1.4)$$

In theory, it is essential that the entire control surface exists out of a wall (a rigid enclosure) to guarantee that the net fluid force is measured accurately, figure 1*b*. In practice, however, viscous shear forces can be ignored if the Reynolds number is much larger than unity, because inertial pressure forces dominate (although shear is not precisely zero except in superfluids (*19*)):

$$\bar{F}(t) \approx -\underbrace{\iint_{CS}p\bar{n}dS}_{pressure} \quad (1.5)$$

Finally, the net acceleration of the moving object or animal can be calculated by dividing the net measured fluid force by the net associated mass:

$$\bar{a}(t) = \bar{F}(t)/M(t) \quad (1.6)$$

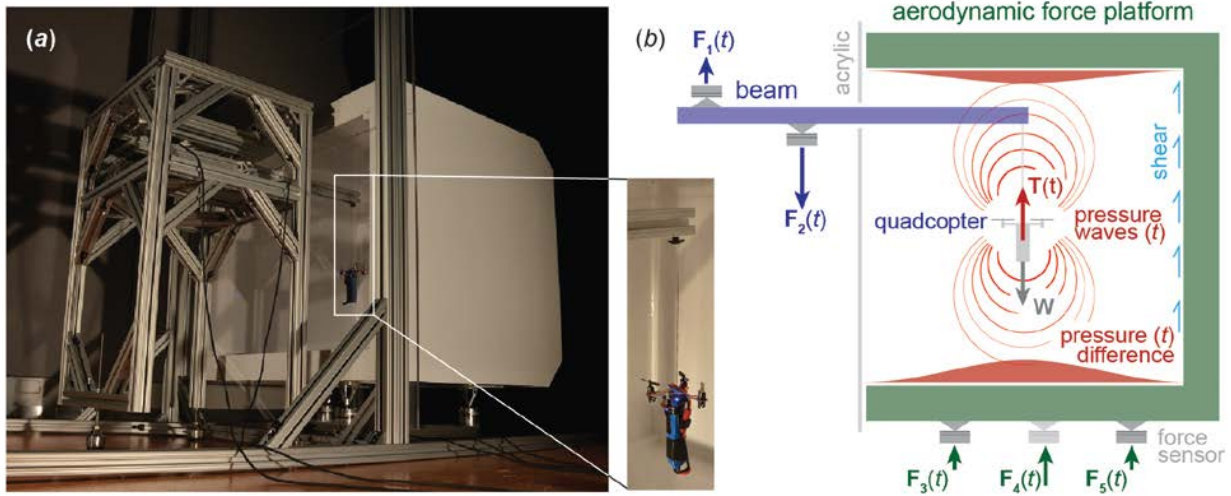

**Figure 1.** Aerodynamic force platform (AFP) working principle. (*a*) Validation of the platform: an overweight quad copter, hung from a beam instrumented with load cells, suspended in the AFP (an instrumented box). The inset shows a close-up of the quad copter and its elongated battery—too heavy to take off. (*b*) Based on Newton's third law, the thrust force of the quad copter, **T**, is balanced by the beam's support forces: $\mathbf{F}_1$–$\mathbf{F}_2$. The thrust force, **T**, is transferred to air, which transfers it as a pressure force normal (and a small shear force tangential) to the walls of the aerodynamic force platform, resulting in the ground reaction force: $\mathbf{F}_3+\mathbf{F}_4+\mathbf{F}_5$. Pressure waves transfer fluctuations in the thrust force at the speed of sound to the surrounding air, and ultimately the platform.

## 3. Experimental validation

We developed an AFP, calibrated it, measured its natural frequency, and validated it using independent load cell measurements on a tethered quadcopter (figure 1*a*; 2*a*). To obtain the net unsteady force with a mechanical implementation of the contour surface integral, the wall needs to be instrumented with load cells that measure the net fluid forces (and moments) that act on the wall. Similar to terrestrial force platforms (*16*), the natural frequency of the instrumented walls of the AFP needs to be an order of magnitude greater than the highest-frequency force-fluctuation of interest (this requires high stiffness and lightweight design). Further, the sensors need sufficient resolution to detect the smallest forces and sufficient dynamic range to resolve the largest forces.

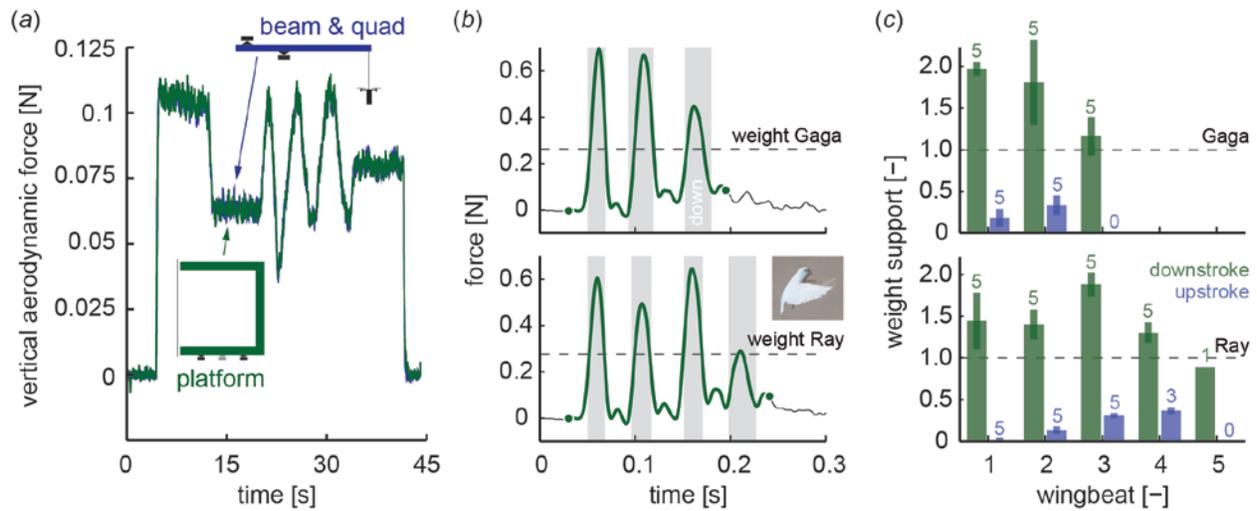

**Figure 2.** The aerodynamic force platform measures weight support of freely flying birds *in vivo*. (*a*) The quadcopter's thrust measured with the platform (green) *vs.* beam (blue) overlap, confirming that the platform is accurate. (*b*) Force-platform measurements of two pacific parrotlets (Gaga & Ray) flying between two perches at 0.28m distance in the AFP (4$^{th}$ order Butterworth filter with 60Hz cut-off; green circle, take-off and landing; gray area, downstroke; photograph of a parrotlet, Linda Cicero, Stanford). (*c*) Calculation of wingbeat-averaged weight support based on raw data (flights, *n*=5; birds, *n*=2).

Our first-generation AFP is essentially a lightweight and stiff instrumented box with one open side for easy access that is covered with an acrylic plate, figure 1*a*. The walls are made out of thin (1mm) balsa wood sheets that are combined into a sandwich structure that maximize platform stiffness with respect to weight (outer height × width × depth; 0.530 × 0.634 × 0.507m, inner; 0.420 × 0.525 × 0.452m). The box is supported by three Nano 43 sensors (6-axis, with SI-9-0.125 calibration, ATI Industrial Automation) that sample force at 1ms intervals with 2 mN resolution. To precisely resolve vertical force, the AFP is connected statically determined (moment free) to the three load cells. The sensors are arranged such that all are equally preloaded by AFP weight (1.79 kg; linear calibration coefficient: 0.989; $r^2$ = 1.000, rounded to three decimals). The natural frequency was measured by popping a balloon five times near the platform. The natural frequency in the vertical (thrust) direction is 132 Hz, which is weakly coupled to a small-amplitude 105 Hz mode.

To validate and evaluate the accuracy of the AFP, we need independent, ground-truth, aerodynamic force measurements. For this we attached a quadcopter (Estes 4606 Proto X Nano, with a 1500 mAh LiPo battery) to an instrumented aluminum beam with a Kevlar tether (two Nano 43 sensors; natural frequency beam + quadcopter, 138 Hz; linear calibration coefficient: 0.997; $r^2$ = 1.000, rounded). Visual comparison of the beam versus AFP measurement suggest that the correspondence in force measurement is remarkably close for a hand-controlled thrust profile, figure 2*a*. To test this systematically, we modified the RC control of the quadcopter to transmit semi-sinusoidal (mean force 80 mN; amplitude 27 mN) and constant thrust profiles (80 mN) using an

Arduino Uno microcontroller. All measured quadcopter thrust profiles were filtered using a fourth-order Butterworth filter with a cutoff frequency of 30 Hz (MATLAB R2010a).

Comparison of the impulse and force ratio of the thrust profiles measured with the AFP versus beam show that our AFP is accurate within 2%—which is equivalent to the load cell resolution limit of 2mN (Table I). Using cross-correlation we find the time delay of the AFP is within its natural vibration period, and thus time resolved. This time delay, due to the transfer function of the AFP, is an order of magnitude larger than the delay due to the speed of sound, showing the delay is primarily determined by the AFP's dynamics. The 2% higher force measured by the AFP is equivalent to the missing shear force that should act on the acrylic plate, which is not instrumented, assuming the Blasius equation for boundary layer friction of a flat plate (*20*). Our back-on-the-envelope estimate of the net shear force, for a measured near-wall flow speed of 0.95 ms$^{-1}$ (with a hotwire) at 25°C, gives a shear force equal to +0.8% of the total force, below the load cells' resolution, but in the right direction. This supports our notion that we can safely ignore shear at Reynolds numbers greater than unity ($Re \approx 27,000$ based on AFP height), which greatly simplifies future AFP design. Comparative thrust measurements using an acrylic front plate with and without a circular gap further demonstrate that shear force can be ignored. These experiments show that the gap can be larger than 20% surface area without affecting measurement accuracy. This enables simple validation experiments and interaction with animals flying in the AFP. Finally, we note that the AFP is a special kind of infrasound microphone, for our AFP we calculated a sensitivity of up to 0.008Pa. We found that high sensitivity requires elimination of all infrasound noise sources in the lab; this includes switching off air conditioners that can increase measurement noise by an order of magnitude. We also found that if the AFP is installed statically determined, it does not require special vibration isolation measures.

**Table I.** Validation of AFP versus beam measurement of integrated impulse and instantaneous force of a quadcopter shows that the AFP is accurate and time-resolved.

| experiment | total impulse ratio [-] | ave. force ratio [-] | delay [ms] |
|---|---|---|---|
| constant ($n = 14$) | 1.016 ± 0.011 | 1.017 ± 0.011 | - |
| 0.125 Hz (84 periods) | 1.014 ± 0.006 | 1.014 ± 0.006 | 2 ± 2 |
| 0.250 Hz (84 periods) | 1.010 ± 0.006 | 1.011 ± 0.006 | 6 ± 1 |
| 0.500 Hz (84 periods) | 1.017 ± 0.004 | 1.018 ± 0.005 | 8 ± 1 |

**Table II.** A hole in one of the sidewalls of the AFP has no effect on vertical impulse and force accuracy (average of five recordings of 6 second each).

| diameter hole [m] | area ratio [-] | total impulse ratio [-] | ave. force ratio [-] |
|---|---|---|---|
| 0 | 0 | 1.019 ± 0.004 | 1.019 ± 0.004 |
| 0.100 | 0.036 | 1.020 ± 0.011 | 1.020 ± 0.011 |
| 0.175 | 0.109 | 1.018 ± 0.003 | 1.018 ± 0.003 |
| 0.250 | 0.223 | 1.021 ± 0.003 | 1.021 ± 0.003 |

## 4. *In vivo* demonstration and outlook

To demonstrate that the AFP can directly measure the aerodynamic force generated by a freely flying animal, we trained two pacific parrotlets (*Forpus coelestis*; 28 gram; 0.2 m wingspan; wingbeat frequency 20 Hz) to fly between two perches in the AFP. To enable training and cueing and rewarding of the bird we made a gap in the acrylic front panel of $0.071 m^2$, which has no measurable effect on accuracy (Table II). The parrotlets were trained using positive reinforcement based on millet seed rewards to fly to a target stick and touch it with its beak (food and water *ad libitum*; cages have enrichment, animals were not sacrificed, all training and experimental procedures were approved by Stanford's Administrative Panel on Laboratory Animal Care). Its aerodynamic weight support was measured *in vivo* within a wingbeat using the AFP, while the start and end of the wingbeat were determined with a synchronized high-speed camera at 1000 fps (Phantom M310), figure 2*b*. The recordings demonstrate that generalist birds, such as the parrotlet, do not aerodynamically support their weight with the upstroke during take-off and landing maneuvers, figure 2*c* (irrespective of interspecific differences in flight style). Instead they generate a vertical force of up to twice their body weight during the downstroke.

     The capability of the AFP to measure aerodynamic force *in vivo* is applicable across taxa and addresses the welfare of experimental animals; it is non-invasive, no-touch, and thus relatively low-stress. Future AFPs can be improved by constructing them using sandwich structures consisting of carbon fiber and Nomex honeycomb. Optical access can be improved using tensioned transparent membranes. The measurement sensitivity can be increased with custom load cells that harness extremely precise capacitive or interferometry based displacement sensors. Ultimately, the three dimensional force vector can be resolved with a fully enclosing AFP. Theoretically, this real-time method should work for many animals, robots, and objects that generate a net force in a fluid. Here we have already demonstrated that aerodynamic force platforms can be used to evaluate the aerodynamic force generation of drones *non-intrusively*, and freely flying birds *in vivo*, with remarkable precision.

**Acknowledgements**

We thank A.K. Stowers for help with electronics and calibration. This research is supported by ONR MURI grant N00014-10-1-0951 and the KACST Center of Excelence for Aeronautics and Astronautics at Stanford. The AFP was developed by DL.